\documentclass[proof]{WileyASNA-v1}

\usepackage{graphicx}
\articletype{Article Type}%

\received{20 Nov. 2020}

\begin{document}

\title{Explaining high-braking indice of magnetars SGR 0501$+$4516 and 1E 2259$+$586 using the double magnetic-dipole model}

\author[1,2,3]{Fang-Zhou Yan}\author[1]{Zhi-Fu Gao*}\author[4]{Wen-Shen Yang}\author[4]{Ai-Jun Dong}
\authormark{Fang-Zhou Yan}

\address[1]{Xinjiang Astronomical Observatory, CAS,150, Science 1-Street, Urumqi, Xinjiang, 830011, China}

\address[2]{Key Laboratory of Radio Astronomy, Chinese Academy of Sciences, West Beijing Road, Nanjing, 210008, China}
\address[3]{University of Chinese Academy of Sciences, No.19, Yuquan Road, Beijing, 100049, China}

\address[4]{Shchool of Physics and Electronic Science, Guizhou Normal University, Guiyang, Guizhou, 550001, China}

\corres{*Zhi-Fu Gao. Xinjiang Astronomical Observatory, CAS,150, Science 1-Street, Urumqi, Xinjiang, 830011, China.\\
 \email{zhifugao@xao.ac.cn}}
 \abstract{In this paper, we attribute high braking indices $n>3$ of two magnetars SGR 0501$+$4516 and 1E 2259$+$586 to the
	decrease in their inclination angles using the double magnetic-dipole model proposed by \cite{Hamil2016}. In
	this model, there are two magnetic moments inside a neutron star, one is generated by the rotation effect of a
	charged sphere, $M_{1}$, and the other is generated by the magnetization of ferromagnetically ordered material, $M_{2}$.
    Our calculations indicate that the magnetic moment $M_{2}$ would evolve towards alignment with the spin axis of the two magnetars, and cause their magnetic inclination angles to decrease. We also define a ratio $\eta=M_{2}/M_{1}$, which reflects the magnetization degree, and find that the values of $\eta$ of the two magnetars are about two-orders of magnitude higher than that of rotationally powered pulsar PSR J1640-4631 with $n=3.15(3)$, assuming that they have the same rate of decrease in their inclination angles.}

\maketitle
\section{Introduction}
The spin-down evolution of pulsars has become an important research hot spot in the field of compact objects\,\citep{Dang2020,Kou2018,Magalhaes2012, Magalhaes2016,Oliveira2018,Wen2020a,Yan2011a,Yan2011b,Yan2018,Zhao2019,Zhao2020}. \cite{Allen1997} proposed that the long-term slowdown of a pulsar's rotation follows a power-law form $\tau_{\rm ext} \propto \Omega^{n}$ with $\tau_{\rm ext}$ being the external torque acting on the crust, $\Omega$ the rotational angular velocity, and $n$ the braking index, and gave an expression
\begin{equation}
	\dot{\Omega}=-K\Omega^{n},
	\label{1}
\end{equation}
where $\dot{\Omega}$ is the derivative of $\Omega$, and $K$ is a positive proportional parameter. The braking index of a pulsar is defined as
\begin{equation}
	n=\frac{\Omega\ddot{\Omega}}{\dot{\Omega}^{2}}=\frac{\nu\ddot{\nu}}{\dot{\nu}^{2}}=2-\frac{P\ddot{P}}{\dot{P}^{2}},
	\label{2}
\end{equation}
where $\ddot{\Omega}$ is the second derivative of $\Omega$, $\nu=\Omega/2\pi$ the spin frequency, and $P=1/\nu$ the spin period. A pure magneto-dipole radiation\,(MDR) model predicts the braking index $n= 3$. However, the observed braking indices of pulsars always deviate from three predicted by the pure MDR model. \cite{Michel1970} proposed the origin of the alignment and slowing down torque with the corresponding equations of motion
\begin{equation}
	I\dot{\alpha}=-\frac{2M^{2}\sin\alpha \cos\alpha \Omega^{2}}{3c^{3}}
	\label{3}
\end{equation}
when pulsars are slowed down, where $I$ is the star's moment of inertia, $\alpha$ is the inclination angle between the rotation and magnetic axes, $\dot{\alpha}$ is the derivative of $\alpha$, $M=B_{d}R^{3}$ is the magnetic moment\,($B_{d}$ is the surface dipole magnetic field strength at the equator, and $R$ the radius of the star), and $c$ is the speed of light in vacuum. In addition, a similar model of the alignment or counteralignment of the magnetic axis of a pulsar with its spin axis was proposed by \,\cite{Macy1974}. A change in the inclination
angle $\alpha$  of a pulsar is usually companied with a change in
pulse profiles \citep{Wen2016a,Wen2016b, Yan2019,Yan2020,Yuan2017,Wen2020b}. A dipole rotating in
vacuum is subject to the radiation field torque \citep{Davis1970}
 \begin{equation}
	I\dot{\Omega}=-\frac{2M^{2}\sin^{2}\alpha\Omega^{3}}{3c^{3}}
	\label{4}
\end{equation}
 The starting point of this work comes from \cite{Michel1970}, \cite{Macy1974} and \cite{Allen1997}. The so-called ``vacuum model'' includes the alignment torque and radiation field torque, described by Eq.\,(3) and
Eq.\,(4), respectively, and allows a change in $\alpha$ and/or a change in the magnetic momentum $M$ of a pulsar.  From
Eq (4), we get the second derivative of $\Omega$,
\begin{eqnarray}
&&\ddot{\Omega}=-\frac{2}{3c^{3}}(3M^{2}\sin^{2}\alpha \Omega^{2}\dot{\Omega}+\nonumber\\
&&~~~~2M\dot{M}\sin^{2}\alpha\Omega^{3}+2M^{2}\sin\alpha\cos\alpha \dot{\alpha}\Omega^{3}).
\label{5}
\end{eqnarray}
Substituting Eqs.(3),(4) and (5) into Eq.\,(2), the observed braking index $n_{\rm obs}$ in vacuum model is expressed as
\begin{equation}
n_{\rm obs}=3+\frac{2\Omega}{\dot{\Omega}}(\frac{\dot{\alpha}}{\tan\alpha}+\frac{\dot{M}}{M}).
\label{6}
\end{equation}
If we ignore the change in $M$, the above equation becomes
\begin{equation}
n_{\rm obs}=3+\frac{2\Omega}{\dot{\Omega}}\frac{\dot{\alpha}}{\tan\alpha}=3+\frac{2\nu}{\dot{\nu}}\frac{\dot{\alpha}}{\tan\alpha}.
\label{7}
\end{equation}
The rotationally powered pulsar PSR J1640$-$4631 has
$\nu$=4.843 Hz and $\dot{\nu}=-2.28\times10^{-11}$\,Hz\,s$^{-1}$
(Archibald et al. 2016), implying a characteristic age of
$\tau_{c}=-\nu/2\dot{\nu}$=3370\,yr and surface dipole field $B_{d}=1.43\times10^{13}I^{1/2}_{45}$\,G, where $I_{45}$ is the
moment of inertia in units of $10^{45}$\,g\,cm$^{2}$\,\citep{Gotthelf2014}. It is the first to have a braking index greater
than 3, $n =3.15(3)$\,(the number in parentheses denotes the error)\,\citep{Archibald2016}, measured with
high precision. \cite{Eksi2016} explained the braking index of PSR J1640-4631
in plasma-filled model and inferred the change in the inclination angle to be at the rate $\dot{\alpha}=-0.23(5)^{\circ}$/century, about 2.5 times smaller in absolute value than the rate $\dot{\alpha}=-0.56^{\circ}$/century of the Crab pulsar through 45 years of observations\,\citep{Lyne2015}.

\setlength{\parindent}{1em}Because of lack of long-term pulsed emission in quiescence and strong timing noise, it is impossible to directly measure the braking index of a magnetar, powered by superhigh multiple magnetic fields. Based on the estimated ages of their potentially associated supernova remnants\,(SNR) and the timing parameter,\cite{Gao2016} measured the mean braking indices of $n=13(4)$, 6.3(1.7) and 32(10) for 1E 1841-045, SGR 0501$+$4516 and 1E 2259$+$586, respectively, using the expression $n=1+P/(\dot{P}t_{SNR})$, where $t_{SNR}$ is the SNR age. Since the uncertainty in estimating the SNR age of 1E 1841-045 is too large, 1E 1841-045 will not be considered in this work. The high braking indices of pulsars have been explained by different models\,(see \cite{Gao2017} for a brief review). We attribute high braking indices $n>3$ of SGR 0501$+$4516 and 1E 2259$+$586 to the decrease in their inclination angles, and estimate their initial magnetic moments, initial inclination angles by using a novel double magnetic-dipole model proposed by\,\cite{Hamil2016}\,(hereafter the HSS model).
\section{The HSS model}
The pulsar's magnetic field was thought to originate from
the following mechanisms: the dynamo effect and the magnetization of a ferromagnetic core in the liquid interior of neutron stars\,(NSs)\,\citep{Sliverstein1969}. Since there is no internal source of energy to generate convection and activate dynamo, it is very difficult to produce a continuous dynamo action in the pulsar. In the HSS model, the pulsar has two dipoles with magnetic moments $M_{1}$ and $M_{2}$, and the distance between two dipole centers is a constant $r$. The second magnetic moment $M_{2}$ is thought to be off-centered dipole, and the inclination angle between the $M_{1}$ and the dipole-dipole axis is $\theta_{1}$. Because the NS interior is in superfluidity phase, the friction received by the rotational region supporting $M_{2}$ is neglected. Due to the interaction between two dipoles, the change rate of $\theta_{2}$ between the $M_{2}$ and the dipole-dipole axis is
\begin{equation}
	\dot{\theta_{2}}=\sqrt{\frac{2M_{1}M_{2}}{I_{2}r^{3}}F(\Theta)},
	\label{8}
\end{equation}
where
$F(\Theta)=(sin\theta_{1i}\sin\theta_{2i}-2\cos\theta_{1i}\cos\theta_{2i})-(\sin\theta_{1f}\sin\theta_{2f}-2\cos\theta_{1f}\cos\theta_{2f})
$, $I_{2}$ is the moment of inertia of the rotating region, $i$ and $f$ represent the initial and final values, respectively.

\setlength{\parindent}{1em}During the motion of the $M_{2}$, its potential energy would convert into the kinetic energy. By taking $C=0$
in Eq.\,(10) of \cite{Hamil2016}, we can obtain the direction of $M_{2}$ when its potential
energy is minimum,
\begin{equation}
	\theta_{2}^{\rm min}=-\arctan(\frac{\tan\theta_{1}}{2}).
	\label{9}
\end{equation}
\setlength{\parindent}{1em}From, Eq.(9), it is obvious that $\theta_{2}^{\rm min}$ should be negative, meaning that the minimum energy position should locate the clockwise direction of the dipole-dipole axis. For an arbitrary initial $\theta_{2}$, the physical law indicates that the $M_{2}$ will freely rotate around the minimum energy position, which is similar to a simple pendulum.

\setlength{\parindent}{1em}According to the geometric relationship, the inclination angle and its change rate are determined by
\begin{equation}
	~~~~~\alpha=\theta_{1}-\theta_{2},~~~{\rm and}~~\dot{\alpha}=-\dot{\theta}_{2},
	\label{10}
\end{equation}
respectively, where $\theta_{1}$ keeps constant, positive (negative) angles $\theta_{1}$ and $\theta_{2}$ are measured from the dipole-dipole axis counterclockwise\,(clockwise), the rotation axis is taken as $\alpha=0$, an increase\,(decrease) in $\alpha$ and a consequent
movement away\,(toward) the axis of rotation.
In this work, for simplicity, we take range of $-90^{\circ}<\theta_{2}<0$ in our simulation.
\section{Applied to two high-$n$ magnetars }
\subsection{Relation between $\dot{\alpha}$ and $\alpha$}
SGR 0501$+$4516 was first discovered by the Swift $\gamma$ -ray observatory, and is associated with SNR HB9 with an age of $t_{\rm SNR}\sim 1.85-5.85$\,kyr. It has the spin period and the period derivative $\nu$=0.1735\,Hz  and $\dot{\nu}=-1.7\times10^{-13}$\,Hz\,s$^{-1}$, implying a surface dipole magnetic field $B_{d}=1.9\times10^{14}I^{1/2}_{45}$\,G\,\citep{Camero2014}. Its braking index is constrained as $6.3(1.7)$.
AXP 1E2259$+$586 lies within the geometric centre of CTB 109 with an age of $t_{SNR}\sim 10-20$ kyr, and has $\nu=0.143$\,Hz, and $\dot{\nu}=0.967\times10^{-14}$\,Hz\,s$^{-1}$, implying a surface dipole magnetic field $B_{d}=0.59\times10^{14}I^{1/2}_{45}$\,G\,\citep{Olausen2014}. Its constrained value of braking index is $n\sim 32(10)$\,\citep{Gao2016}. Regularly monitored with RXTE from 1996 to 2018, influenced by frequent glitches and outbursts, SGR 0501$+$4516 and
1E 2259$+$586 exhibited large fluctuations in spin-down parameters $\nu$ and $\dot{\nu}$. These large
decreases in $\dot{\nu}$ were explained as the effects of timing-noise torque
and outbursts, instead of as a secular dipole magnetic field decay\,(e.g.\cite{Lin2020}. Their outbursts were explained as resulting from internal multiple magnetic field decay\,\citep{Duncan1992}.

According to the HSS model, when $M_{2}$ is rotating in the vicinity of the equilibrium
position with the counter clockwise direction, $\dot{\theta}_{2}> 0$, hence
$\dot{\alpha}<0$, we obtain a braking index greater than three from Eq.\,(7), which can naturally explain high braking indices of SGR 0501$+$4516 and 1E 2259$+$586. Noted that, if $\dot{\theta}_{2}<0$, it will
result in a braking index smaller than three\, however, this case is not the subject of this work.
From Eq.\,(7), the change rate of the inclination angel, $\dot{\alpha}$, is estimated from
\begin{equation}
	\dot{\alpha}=(n_{\rm obs}-3)\frac{\dot{\Omega}}{2\Omega}\tan\alpha=(n_{\rm obs}-3)\frac{\dot{\nu}}{2\nu}\tan\alpha.
	\label{11}
\end{equation}

\setlength{\parindent}{1em}Inserting the values of $\nu$ and $\dot{\nu}$ into Eq.\,(11), we make the plots of $-d\alpha/dt$ versus $\alpha$ for the two high-$n$ magnetars. In Fig.\,1, we take $n_{\rm obs}\approx \bar{n}=6.3$ and 32 for SGR 0501$+$4516 and 1E 2259$+$586, respectively. Obviously, $-d\alpha/dt$ increases with the increase in $\alpha$ when $n_{\rm obs}$ is given.
\begin{figure}[t]
	\centerline{\includegraphics[height=13pc]{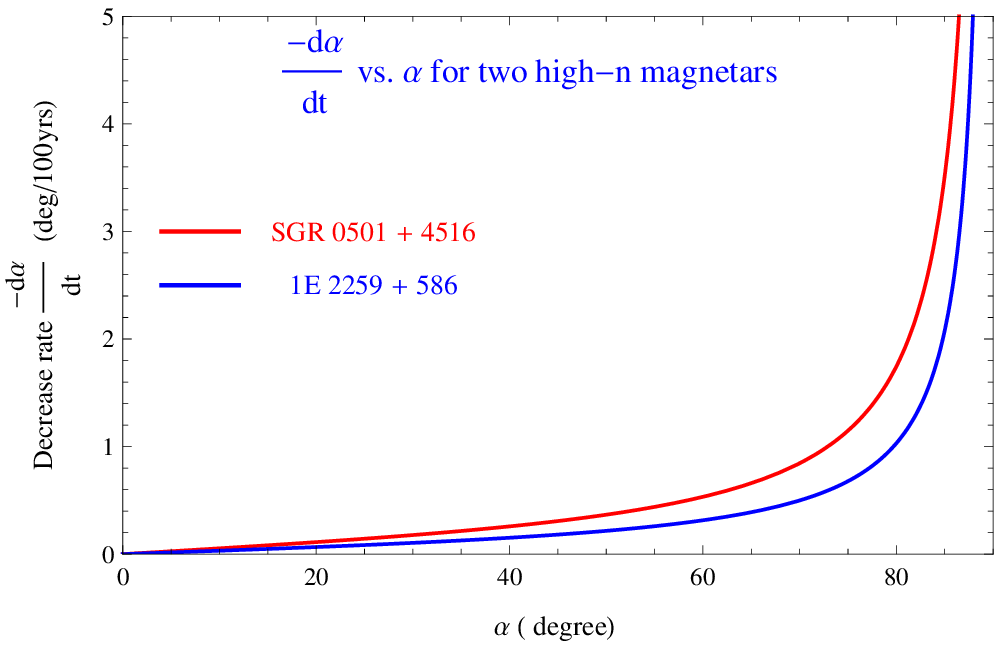}}
	\caption{Relations of $\dot{\alpha}$ and $\alpha$ of two magnetars\label{fig1}}
	\vspace{-0.7cm}
\end{figure}
To date, due to the lack of polarization observations, we cannot obtain the actual value of $\alpha$ of a magnetar and its change rate $\dot{\alpha}$ observationally. In this work we assume that SGR 0501$+$4516 and 1E 2259$+$586 are experiencing a decrease in $\alpha$, it is possible that the actual values of $\dot{\alpha}$ of the two magnetars are near to that of PSR J1640-4631\,\citep{Eksi2016} or to that of the Crab pulsar\,\citep{Lyne2015}. We can obtain the information related to the inclination angles, their decrease rates and magnetic moments of the two magnetars by numerically simulating.
\subsection{SGR 0501 and 1E 2259}
If the current decrease rate of $\alpha$ of SGR 0501$+$4516 is same to that of PSR J1640-4631, $\dot{\alpha}=-0.23^{\circ}$/century\,\citep{Eksi2016}, since its average age is $t=t_{\rm SNR}=3.85$\,kyr, we take a mean decrease rate $\bar{\dot{\alpha}}=-0.115^{\circ}$/century, the total decrease of $\alpha$ is then estimated as $\Delta\alpha= 38.5\times(-0.115^{\circ})\approx 4^{\circ}$. Inserting $\nu$ and $\dot{\nu}$ into Eq.\,(7). we get the current value of $\alpha$, $\alpha_{f}\approx37^{\circ}$, and the initial value of $\alpha$, $\alpha_{i}=41^{\circ}$. In order to obtain a negative $\dot{\alpha}$, $\theta_{2}$ should be less than $\theta_{2}^{\rm min}$, and $\theta_{1}$ should be less than a critical angle $\theta_{1}^{\rm crit}$, which is determined by
\begin{equation}
\theta_{1}^{\rm crit}\leq\alpha_{f}-\arctan(\frac{\tan(\theta_{1}^{\rm crit})}{2}).
\label{12}
\end{equation}

\setlength{\parindent}{1em}From Eq.\,(12), we get $\theta_{1}^{\rm crit}=24.4^{\circ}$ and $\alpha_{f}=37^{\circ}$.
If the current decrease rate of $\alpha$ of the star is same to that of the Crab, we obtain $\Delta\alpha\approx 11^{\circ}$, $\alpha_{f}\approx61^{\circ}$, and $\alpha_{i}\approx 72^{\circ}$ by taking a mean decrease rate $\bar{\dot{\alpha}}=-0.28^{\circ}$/century. In the same way, we obtain  $\theta_{1}^{\rm crit}=39.1^{\circ}$, corresponding to $\alpha_{f}=61^{\circ}$.
By choosing particular values of $\theta_{1}$, we calculate the values of $F(\Theta)$ and $\theta_{2}^{\rm min}$.
In Table 1, Case A and Case B correspond to $\bar{\dot{\alpha}}=-0.115^{\circ}$/century and $-0.28^{\circ}$/century, respectively, which are same to those of PSR J1640-4631 and the Crab, respectively. It's obvious that $F(\Theta)$ decreases with the increase in $\theta_{1}$. Comparing $\theta_{2}$ with $\theta_{2}^{\rm min}$, the $M_{2}$ should rotate along the counter clockwise, and the angle $\alpha$ decreases.

We choose a medium-mass NS with $m=1.45M_{\odot}$ and $R=11.5$\,km, corresponding to the moment of inertia $I= 1.534(1)\times
10^{45}$\,g\,cm$^{2}$ in the BSK26 EoS. The moment of inertia of $M_{1}$ is $I_{1}\approx I$. Taking $I_{2}$ as $1\%$ of the total $I$ and $r=0.5R$, the magnetic moment $M_{2}$ of SGR 0501+4516 is calculated as $M_{2}=B_{d}R^{3}\approx 3.58\times10^{32}$\,G\,cm$^{3}$. If $\dot{\theta_{2}}=-\dot{\alpha}=0.23^{\circ}$/century, taking $\theta_{i}=0^{\circ}, 10^{\circ}, 20^{\circ}$ and $24^{\circ}$, respectively, and using Eq.(8), we calculate the values of $M_{1}$ and $M_{2}/M_{1}$ and list them in Table 2. If $\dot{\theta_{2}}=-\dot{\alpha}=0.56^{\circ}$/century, taking $\theta_{i}=0^{\circ}, 10^{\circ}, 20^{\circ}, 30^{\circ}$ and $38^{\circ}$, respectively, we also calculate the values of $M_{1}$ and $M_{2}/M_{1}$ and list them in Table 2.

\makeatletter\def\@captype{table}\makeatother
\captionof{table}{\color{Black}Main angle parameters of SGR 0501+4516 in the HSS model  }
\begin{tabular}{p{0.496cm}p{0.5cm}p{0.5cm}p{1.004cm}p{0.769cm}p{1.1cm}c}
	\toprule
	\textbf{$\alpha_{f}$}&\textbf{$\theta_{1i}$}&\textbf{$\theta_{1f}$}&\textbf{$\theta_{2i}$}&\textbf{$\theta_{2f}$}&\textbf{$\theta_{2}^{\rm min}$}&
	\textbf{$F(\Theta)$}\\
	\bottomrule
	& & & Case A& &  &\\
	\midrule
	$37^{\circ}$ & $0^{\circ}$  & $0^{\circ}$  &$-41^{\circ}$  &$-37^{\circ}$ & $0^{\circ}$& $0.0878$\\
	$37^{\circ}$ & $5^{\circ} $ & $5^{\circ}$  &$-36^{\circ} $ &$-32^{\circ}$ & $-2.5^{\circ}$&$0.0727$\\
	$37^{\circ}$ & $10^{\circ}$ & $10^{\circ}$ &$-31^{\circ}$  &$-27^{\circ}$ & $-5.0^{\circ}$ &$0.0561$\\
	$37^{\circ}$ & $15^{\circ}$ & $15^{\circ}$ &$-26^{\circ}$  &$-22^{\circ}$ & $-7.6^{\circ}$ &$0.0383$\\
	$37^{\circ}$ & $20^{\circ}$ & $20^{\circ}$ &$-21^{\circ}$  &$-17^{\circ}$ & $-10.3^{\circ}$ &$0.0201$\\
	$37^{\circ}$ & $24^{\circ}$ & $24^{\circ}_{\ast}$ &$-17^{\circ}$  &$-13^{\circ}$ & $-12.6^{\circ}$&$0.0056$\\
	\bottomrule
	& & & Case B & &  & \\
	\midrule
	$61^{\circ}$ & $0^{\circ}$  & $0^{\circ}$  &$-72^{\circ}$  &$-61^{\circ}$ & $0^{\circ}$& $0.3516$\\
	$61^{\circ}$ & $5^{\circ} $ & $5^{\circ}$  &$-67^{\circ} $ &$-56^{\circ}$ & $-2.5^{\circ}$&$0.3277$\\
	$61^{\circ}$ & $10^{\circ}$ & $10^{\circ}$ &$-62^{\circ}$  &$-51^{\circ}$ & $-5.0^{\circ}$ &$0.2965$\\
	$61^{\circ}$ & $15^{\circ}$ & $15^{\circ}$ &$-57^{\circ}$  &$-46^{\circ}$ & $-7.6^{\circ}$ &$0.2589$\\
	$61^{\circ}$ & $20^{\circ}$ & $20^{\circ}$ &$-52^{\circ}$  &$-41^{\circ}$ & $-10.3^{\circ}$ &$0.2162$\\
	$61^{\circ}$ & $25^{\circ}$ & $25^{\circ}$ &$-47^{\circ}$  &$-36^{\circ}$ & $-13.1^{\circ}$&$0.1692$\\
	$61^{\circ}$ & $30^{\circ}$  & $30^{\circ}$  &$-42^{\circ}$  &$-31^{\circ}$ & $-16.1^{\circ}$& $0.1204$\\
	$61^{\circ}$ & $35^{\circ} $ & $35^{\circ}$  &$-37^{\circ} $ &$-26^{\circ}$ & $-19.3^{\circ}$&$0.0703$\\
	$61^{\circ}$ & $38^{\circ}$ & $38^{\circ}$ &$-34^{\circ}$  &$-23^{\circ}$ & $-21.3^{\circ}$ &$0.0404$\\
	\bottomrule
	\vspace{-0.2cm}
\end{tabular}
\makeatletter\def\@captype{table}\makeatother
\vspace{-0.2cm}

\vspace{-0.1cm}
\captionof{table}{\color{Black}Partial values of $\alpha_{f}$, $F(\Theta)$, $M_{1}$ and $M_{2}$ and $\eta$ for SGR 0501+4516 }
\begin{tabular}{p{0.50cm}p{0.95cm}p{1.6cm}p{1.6cm}p{1.6cm}}
	\toprule
	$\alpha_{f}$  &$F(\Theta)$ & $M_{2}$ & $M_{1}$ &$\eta$ \\
	\midrule
	&   &  G\,cm$^{3}$ & G\,cm$^{3}$& \\
	\midrule
	$37^{\circ}$ &0.0878 &   $3.58\times10^{32}$ &$7.52\times10^{4}$ & $4.76\times10^{27}$\\
	$37^{\circ}$ &0.0561 &   $3.58\times10^{32}$ &$1.18\times10^{5}$ & $3.04\times10^{27}$\\
	$37^{\circ}$ &0.0201 &   $3.58\times10^{32}$&$3.28\times10^{5}$ & $1.09\times10^{27}$\\
	$37^{\circ}$ &0.0056 &   $3.58\times10^{32}$&$1.18\times10^{6}$ & $3.04\times10^{26}$\\
	\midrule
	$61^{\circ} $&0.3516 &   $3.58\times10^{32}$ &$1.11\times10^{5}$   &$3.22\times10^{27}$\\
	$61^{\circ} $& 0.2965&   $3.58\times10^{32}$ &$1.32\times10^{5}$  &$2.71\times10^{27}$\\
	$61^{\circ} $ & 0.2162 &    $3.58\times10^{32}$&$1.81\times10^{5}$ &$1.98\times10^{27}$\\
	$61^{\circ} $ & 0.1204&   $3.58\times10^{32}$&$3.25\times10^{5}$   &$1.10\times10^{27}$\\
	$61^{\circ} $ & 0.0404&   $3.58\times10^{32}$&$9.68\times10^{5}$   &$3.69\times10^{26}$\\
	\bottomrule
\end{tabular}
\vspace{0.3cm}

\setlength{\parindent}{1em}In order to explore the information of paramagnetic magnetization of a NS in its early ferromagnetic phase, we define a model-dependent ratio, $\eta=M_{2}/M_{1}$, which reflects the magnetization degree and depends on angle parameters, e.g., $\dot{\alpha}$ and $F(\Theta)$. From Table 2, we find that the values of $\eta$ are about $10^{26}-10^{27}$ for SGR 0501+4516. Fig.\,2 denotes $n$ as a function of $\alpha$ for SGR 0501+4516, the red and blue solid lines are for $\dot{\alpha}=-0.23^{\circ}$/century, $\dot{\alpha}=-0.56^{\circ}$/century, respectively, and the horizontal blue dotted line and the surrounding shaded region denote, respectively, the constrained value of $n=13$ and its possible range given by the uncertainty 4 of SGR 0501$+$4516. From Fig. 2, it is
obvious that $n$ decreases with the increase in $\alpha$ when the values of $\dot{\alpha}$, $\nu$ and $\dot{\nu}$ are given. If the change
rate of $\alpha$ of SGR 0501$+$4516 is similar to the Crab, it probably has a larger inclination angle.
\begin{figure}[t]
	\centerline{\includegraphics[height=13pc]{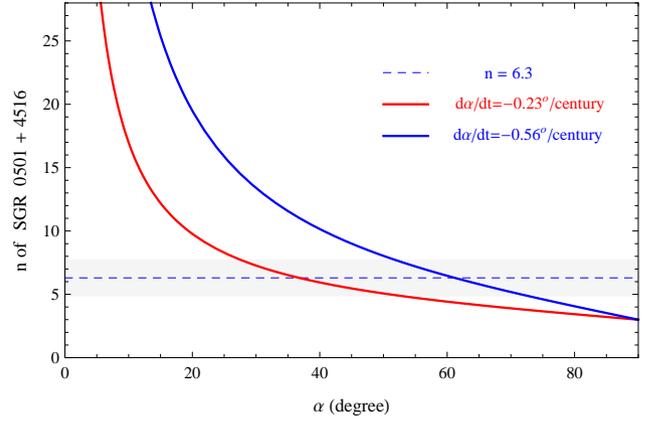}}
	\caption{Relation of $n$ and $\alpha$ for SGR 0501+4516.\label{fig2}}
	\vspace{-0.8cm}
\end{figure}
If the current change rate of $\alpha$ for 1E 2259$+$586 is same to that of PSR J1640-4631, $\dot{\alpha}=-0.23^{\circ}$/century\,\citep{Eksi2016}. Since its SNR age is $t\approx t_{\rm SNR}$=15\,kyr, assuming a mean change rate $\bar{\dot{\alpha}}=-0.115^{\circ}$/century ,the total change of $\alpha$ is $\Delta\alpha=-150\times0.115^{\circ}/\approx17^{\circ}$. Inserting $\nu$ and $\dot{\nu}$ into Eq.(7), we obtains the current value $\alpha_{f}\approx51^{\circ}$, and the initial value of $\alpha$, $\alpha_{i}=68^{\circ}$. From Eq.(12),the critical angle $\theta_{1}^{crit}=33.1^{\circ}$. If the current change rate of $\alpha$ is same to that of the Crab, $\dot{\alpha}=-0.56^{\circ}$/century, in the same method, the current value of $\alpha_{f}=72^{\circ}$, but the initial angle $\alpha_{i}>90^{\circ}$, which is unreasonable. Thus, the change rate of $\alpha$ for 1E 2259$+$586 could be smaller than that of the Crab pulsar in the HSS model.

\setlength{\parindent}{1em}Assuming $\bar{\dot{\alpha}}=-0.115^{\circ}$/century and choosing several particular angles of $\theta_{1}$, we calculate the values of $F(\Theta)$ and $\theta_{2}^{\rm min}$ for 1E 2259$+$586. From Table 3, $F(\Theta)$ decreases with the increase in $\theta_{1}$. Comparing $\theta_{2}$ with $\theta_{2}^{\rm min}$, the $M_{2}$ of the star should also rotate along the counter clockwise. Using the same EoS, the magnetic moment $M_{2}$ of 1E 2259$+$586 is calculated as $M_{2}=B_{d}R^{3}\approx 1.12\times10^{32}$\,G\,cm$^{3}$.  We calculate the values of $M_{1}$, $M_{2}$ and $\eta$ and list them in Table 4. It is found that both $F(\Theta)$ and $\eta$ decrease with the increase in $\theta_{1i}$, and the values of $\eta$ are also in the range of $10^{26}-10^{27}$. Fig.\,3 denotes $n$ as a function of $\alpha$ for 1E 2259$+$586 assuming $\dot{\alpha}=-0.23^{\circ}$/century. As a comparison, using the same method, we calculate the values of $\dot{\theta}_{2}$, $F(\Theta)$, $M_{1}$ and $M_{2}$ and $\eta$ for PSR J1640$-$4631 and list them in Table 5. We find that the values of $M_{2}$ of PSR J1640$-$4631 are about 4-10 times smaller than those of 1E 2259$+$586 and SGR 0501$+$4516, and the values of $\eta$ of PSR J1640$-$4631 are 1-2 orders of magnitude less than those of the two magnetars. Through PSR J1640$-$4631 is a rotationally powered X/$\gamma$-ray pulsar, it posses a higher dipole magnetic field $\sim 10^{13}$\,G and lacks of radio emission. These features are close to magnetars. However, for a common radio pulsar with typical dipole field $B_{d}^{'}\sim10^{12}$\,G, its ratio of $\eta^{'}$ could be 3-5 orders of magnitude less than that of a canonical magnetar with $B_{d}\sim10^{14}-10^{15}$\,G.

\vspace{-0.1cm}
\makeatletter\def\@captype{table}\makeatother
\captionof{table}{\color{Black}Main angle parameters of 1E 2259+586}
\begin{tabular}{lllllll}
	\bottomrule
	$\alpha_{f}$& $\theta_{1i}$& $\theta_{1f}$ &$\theta_{2i}$ &$\theta_{2f}$&$\theta_{2}^{\rm min}$&$F(\Theta)$\\
	\midrule
	$51^{\circ}$ & $0^{\circ}$  & $0^{\circ}$  &$-68^{\circ}$  &$-51^{\circ}$ & $0^{\circ}$& $0.5094$\\
	$51^{\circ}$ & $5^{\circ} $ & $5^{\circ}$  &$-63^{\circ} $ &$-46^{\circ}$ & $-2.5^{\circ}$&$0.4645$\\
	$51^{\circ}$ & $10^{\circ}$ & $10^{\circ}$ &$-58^{\circ}$  &$-41^{\circ}$ & $-5.0^{\circ}$ &$0.4096$\\
	$51^{\circ}$ & $15^{\circ}$ & $15^{\circ}$ &$-53^{\circ}$  &$-36^{\circ}$ & $-7.6^{\circ}$ &$0.3457$\\
	$51^{\circ}$ & $20^{\circ}$ & $20^{\circ}$ &$-48^{\circ}$  &$-31^{\circ}$ & $-10.3^{\circ}$ &$0.2754$\\
	$51^{\circ}$ & $25^{\circ}$ & $25^{\circ}$ &$-43^{\circ}$  &$-26^{\circ}$ & $-13.1^{\circ}$&$0.2005$\\
	$51^{\circ}$ & $32^{\circ}$ & $32^{\circ}$ &$-36^{\circ}$  &$-19^{\circ}$ & $-17.4^{\circ}$&$0.0925$\\
	\bottomrule
	\vspace{-0.1cm}
\end{tabular}
\begin{footnotesize}
\end{footnotesize}

\makeatletter\def\@captype{table}\makeatother
\vspace{-0.1cm}
\captionof{table}{\color{Black} Partial values of $\alpha_{f}$, $F(\Theta)$, $M_{1}$, $M_{2}$ and $\eta$ for 1E 2259+586}
\begin{tabular}{p{0.50cm}p{0.95cm}p{1.7cm}p{1.408cm}p{1.58cm}}
	\toprule
	$\alpha_{f}$ &$F(\Theta)$ &$M_{2}$ &$M_{1}$ &$\eta$\\
	\midrule
	&            &G\,cm$^{3}$ & G\,cm$^{3}$ & \\
	\midrule
	$51^{\circ}$ &0.5094   &   $1.12\times10^{32}$&$4.14\times10^{4}$ &$2.70\times10^{27}$\\
	$51^{\circ}$ &0.4645   &   $1.12\times10^{32}$&$4.54\times10^{4}$ &$2.47\times10^{27}$\\
	$51^{\circ}$ &0.4096   &   $1.12\times10^{32}$&$5.15\times10^{4}$ &$2.17\times10^{27}$\\
	$51^{\circ}$ & 0.3457  &   $1.12\times10^{32}$&$4.71\times10^{4}$ &$3.09\times10^{27}$\\
	$51^{\circ}$ & 0.2754  &   $1.12\times10^{32}$&$6.10\times10^{4}$ &$1.83\times10^{27}$\\
	$51^{\circ}$ & 0.2005  &   $1.12\times10^{32}$&$1.05\times10^{5}$ &$1.06\times10^{27}$\\
	$51^{\circ}$ & 0.0925  &   $1.12\times10^{32}$&$2.28\times10^{5}$ &$4.91\times10^{26}$\\
	\bottomrule
	\vspace{0.2cm}
\end{tabular}
\begin{figure}[t]
	\centerline{\includegraphics[height=13pc]{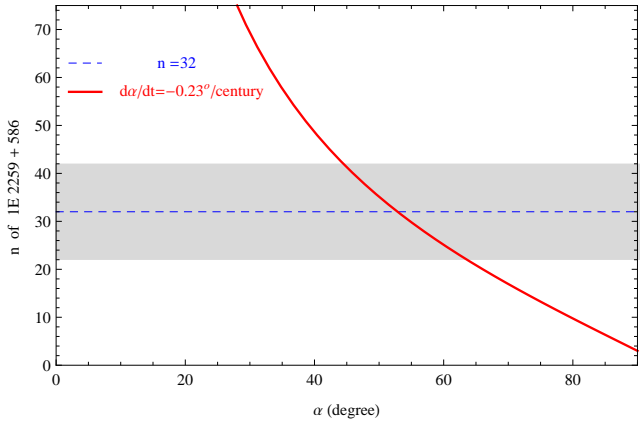}}
	\caption{Relation of $n$ and $\alpha$ for 1E 2259+586.\label{fig3}}
	\vspace{-0.8cm}
\end{figure}

\setlength{\parindent}{1em} A simple estimation is given as follows. Firstly, from Eq.\,(8), we get $M_{2}/M_{1}=2M^{2}_{2}F(\Theta)/(\dot{\alpha}^{2}I_{2}r^{3})$ in the HSS model. The differences in EoSs of a common radio pulsar and a magnetar may be very small, and can be ignored. Assuming a same decrease rate $\dot{\alpha}$, we give an approximation expression
\begin{equation}
\frac{\eta}{\eta^{'}}\approx\frac{M^{2}_{2}F(\Theta)}{M^{'2}_{2}F^{'}(\Theta)}\approx\frac{B^{2}_{d}F(\Theta)}{B^{'2}_{d}F^{'}(\Theta)},
\label{13}
\end{equation}
where $\eta$ and $\eta^{'}$ are for a magnetar and a common radio pulsar, respectively.

\vspace{0.2cm}
\makeatletter\def\@captype{table}\makeatother
\captionof{table}{\color{Black} Partial values of $\dot{\theta}_{2}$,  $F(\Theta)$, $M_{1}$ and $M_{2}$ and $\eta$  for PSR J1640-4631}
\begin{tabular}{p{0.88cm}p{0.88cm}p{1.52cm}p{1.408cm}p{1.52cm}}
	\toprule
	$\dot{\theta}_{2}$  &$F(\Theta)$ &$M_{2}$ &$M_{1}$ & $\eta$\\
	\midrule
	$^{\circ}$/cent   &            &G$\cdot$cm$^{3}$ &G$\cdot$cm$^{3}$& \\
	\midrule
	0.23 &0.122  &   $2.7\times10^{31}$&$7.2\times10^{5}$ &$3.7\times10^{25}$\\
	0.23 &0.097  &   $2.7\times10^{31}$&$9.1\times10^{5}$ &$2.9\times10^{25}$\\
	0.56 & 0.657  &   $2.7\times10^{31} $&$7.9\times10^{5}$ &$3.4\times10^{25}$\\
	0.56  & 0.919  &   $2.7\times10^{31}$&$5.7\times10^{5}$ &$4.8\times10^{25}$\\
	\bottomrule
\end{tabular}
\vspace{0.4cm}

\setlength{\parindent}{1em}From Eq.\,(13), it is obvious that the degree of magnetization of a magnetar is far higher that that of a common radio pulsar. The above equation implies that there is a difference in the magnetic field formation mechanism\,(including paramagnetic magnetization) between  rotationally powered pulsars and magnetic-field powered magnetars during their early ferromagnetic phases. However, the cause of such a difference is not yet known to date. I hope that this work will be helpful in understanding the nature of magnatars.
\section{Summary}
We have applied a toy two-dipole model to the two high-$n$ magnetars SGR0501$+$4516 and 1E 2259$+$586, and ignored other factors influencing their braking indices, which are attributed to the decrease in inclination angles. For these two sources, their initial magnetic moments, initial inclination angles and average decrease rates of inclination angles are estimated and comparisons with rotationally powered pulsars are presented. Our results could be useful in understanding the nature of magnetars, and will be tested by future observations.
\subsection*{Acknowledgements}
This work was supported by the NSFC (11573009,  U1831120,  U1731238), the Foundation of Science and Technology of Guizhou  Provice([2019]1241,[2017]7349, [2016]4008, (2017) 5726-37)), Outstanding Postdoctoral Foundation of XinJiang and the Foudation of Guizhou Provincial Education Department ((2020)003) and Doctor Starting Up Foundation of Guizhou Normal University (0516134).
\bibliography{FZYan}
\end{document}